\documentclass[aps, pra, preprint, superscriptaddress]{revtex4}

\usepackage{graphicx}
\usepackage{dcolumn}
\usepackage{bm}
\usepackage{epstopdf}

\begin{document}

\title{Inhomogeneity in the ultrafast insulator-to-metal transition dynamics of VO$_2$}

\author{Brian T. O'Callahan}
\thanks{These authors contributed equally to this work.}
\author{Joanna M. Atkin}
\thanks{These authors contributed equally to this work.}
\author{Andrew C. Jones}
\affiliation{Department of Physics, Department of Chemistry, and JILA, University of Colorado, Boulder, CO 80309}
\author{Jae Hyung Park}
\author{David Cobden}
\affiliation{Department of Physics, University of Washington, Seattle, WA 98195}
\author{Markus B. Raschke\footnote{Corresponding author: markus.raschke@colorado.edu}}
\affiliation{Department of Physics, Department of Chemistry, and JILA, University of Colorado, Boulder, CO 80309}

\date{\today}

\begin{abstract}
The insulator-to-metal transition (IMT) of the simple binary compound of vanadium dioxide VO$_2$ at $\sim 340$ K has been puzzling since its discovery more than five decades ago. 
A wide variety of photon and electron probes 
have been applied in search of a satisfactory microscopic mechanistic explanation.
However, many of the conclusions drawn have implicitly assumed a {\em homogeneous} material response. 
Here, we reveal inherently {\em inhomogeneous} behavior in the study of the dynamics of individual VO$_2$ micro-crystals using a combination of femtosecond pump-probe microscopy with nano-IR imaging.
The time scales of the photoinduced bandgap reorganization in the ultrafast IMT vary from $\simeq 40 \pm 8$ fs, i.e., shorter than a suggested  phonon bottleneck, to $\sim 200\pm20$ fs, with an average value of $80 \pm 25$ fs, similar to results from previous studies on polycrystalline thin films. 
The variation is uncorrelated with crystal size, orientation, transition temperature, and initial insulating phase. 
This together with details of the nano-domain behavior during the thermally-induced IMT 
suggests a significant sensitivity to local variations in, e.g., doping, defects, and strain of the microcrystals.
The combination of results points to an electronic mechanism dominating the photoinduced IMT in VO$_2$, but also highlights the difficulty of deducing mechanistic information where the intrinsic response in correlated matter may not yet have been reached.

\end{abstract}

\maketitle
\section{Introduction}
Vanadium dioxide (VO$_2$) is one of the prototypical correlated-electron materials, exhibiting an insulator-metal transition (IMT), with a change in resistivity of several orders of magnitude that can be induced thermally  at $T \sim 340$ K \cite{Morin1959, Imada1998}, or optically \cite{Cavalleri2001}. 
The photoinduced IMT occurs on sub-picosecond timescales, at rates too fast for complete thermalization, and therefore has a non-thermal basis.
This ultrafast transition has been studied using a wide variety of spectroscopies, with short-pulse optical \cite{Becker1994, Cavalleri2001, Cavalleri2004b, Wall2013}, X-ray diffraction \cite{Cavalleri2005,Cavalleri2004,Hada2010}, terahertz \cite{Kubler2007,Nakajima2008,Xue2013}, and electron diffraction \cite{Grinolds2006,Lobastov2007,Baum2007} techniques.
These studies have addressed the ultrafast electron dynamics and lattice structural processes
 which occur during the transition, in additional to slower behavior over multi-picosecond to nanosecond time scales.
However, the mechanism underlying both the thermal and photoinduced insulator-to-metal transition remains unclear, and results conflict with both Mott or Peierls explanations \cite{Mott1968, Zylbersztejn1975, Qazilbash2007,Kubler2007, Tao2012}. 
For example, degenerate pump-probe studies \cite{Cavalleri2004b} revealed a limiting transition timescale of 75 fs, suggesting a phonon bottleneck and therefore a structurally-limited transition. 
In contrast, the observation of coherent phonon oscillations above the apparent threshold for triggering the photoinduced phase transition \cite{Kubler2007} indicates that the metallic phase appears after one V-V phonon oscillation cycle, even though the lattice is still presumably far from equilibrium. 
This suggests that the photoinduced IMT is decoupled from the structural transition.

Much of the work on VO$_2$ has focused on polycrystalline thin films, grown by a variety of techniques \cite{Warwick2014}.
Recently, differences observed in the ultrafast and thermal properties due to anisotropy and grain size in polycrystalline and epitaxial thin film samples \cite{Lysenko2013,Liu2014,Xue2013,Huffman2013} suggest that growth conditions can substantially modify the measured response.
These results contributed to the confusion in the interpretation of previous measurements,
which is further amplified by the use of ultrafast techniques that average over multiple crystallites, subject to large inhomogeneous strain, which could, for example, create mixtures of the different insulating phases. 

In order to access the {\em homogeneous} response, 
we investigate individual VO$_2$ single micro-crystals.
We perform both degenerate pump-probe microscopy (Fig.~1a, for details see Methods) to monitor the femtosecond dynamics following the ultrafast photoinduced excitation, and infrared scattering-scanning near-field optical microscopy ($s$-SNOM) to probe the details of the evolution of the spatial phase competition process in the thermal IMT.
\begin{figure}
\includegraphics[width = \columnwidth]{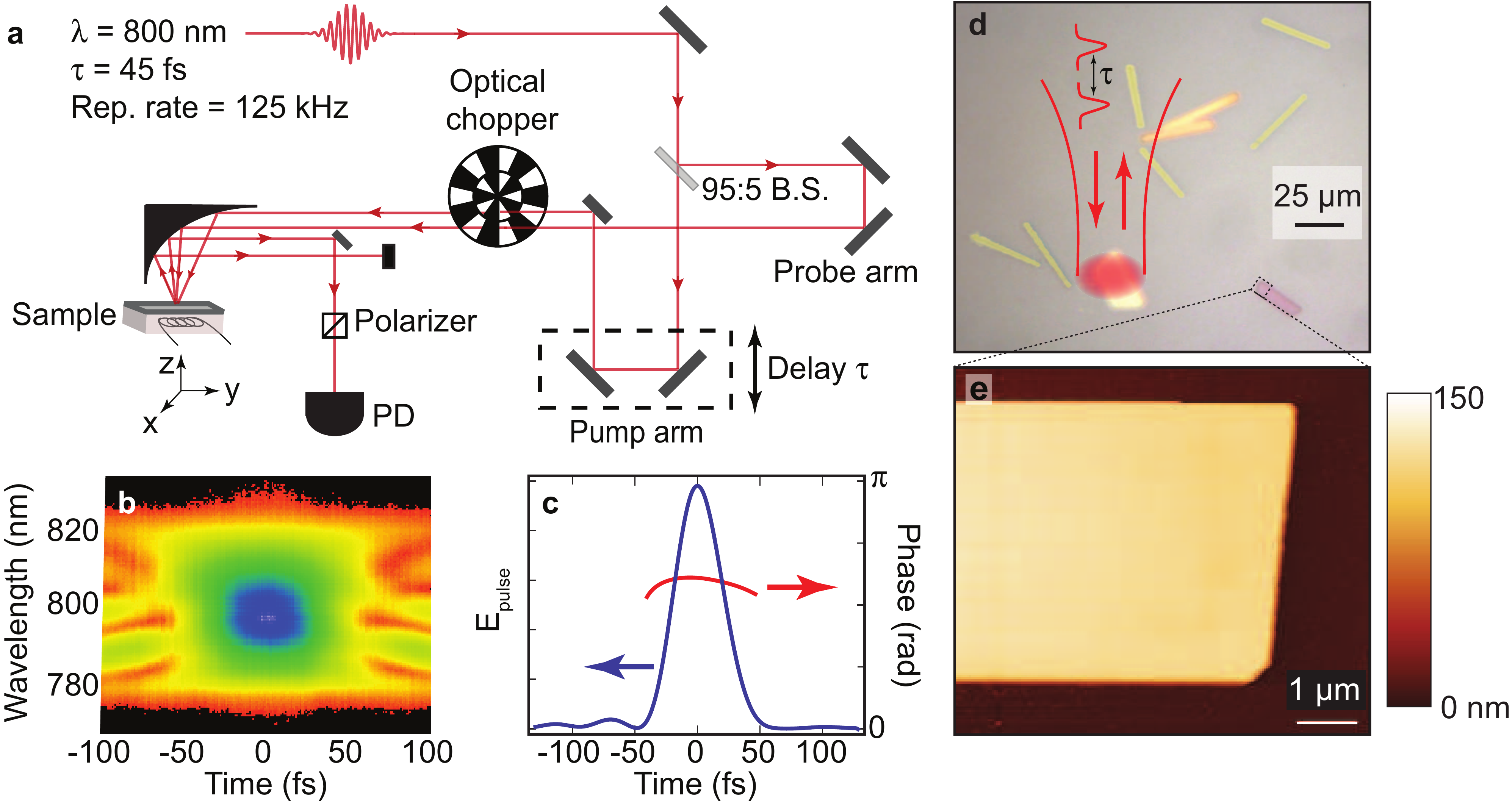}
\caption{Schematic of optical layout for degenerate 800 nm pump-probe (a). 
Typical FROG spectrogram of pulse (b), with full phase and amplitude reconstruction of pulse in time domain (c), showing an approximately 45 fs pulse with small chirp.
Optical microscope image of VO$_2$ microcrystals on Si/SiO$_2$ substrate (d). 
AFM topography (e) showing the homogeneous nature of the microcrystals. 
}
\label{Fig:Pumpprobe}
\end{figure}

\begin{figure}
\includegraphics[width = 12 cm]{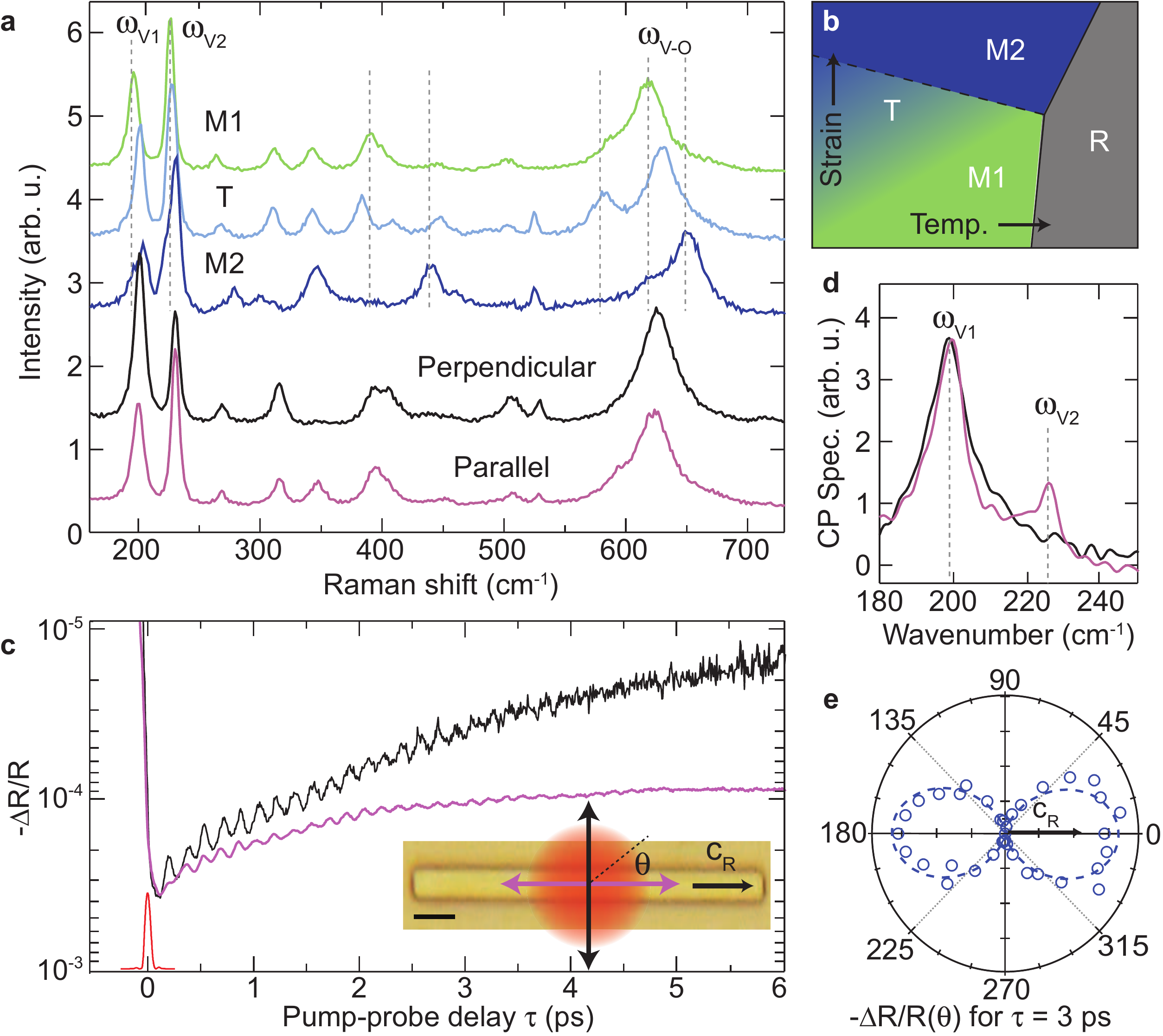}
\caption{Raman spectra of the three insulating structures, M1 (green, microcrystal 1), M2 (dark blue, microcrystal 20), and intermediate triclinic T (light blue, microcrystal 4), allowing full characterization of the initial phase of individual single crystals (a).
Perpendicular (black) and parallel (purple) polarization with respect to $c_R$ a microcrystal in the monoclinic M1 phase.
Strain - temperature diagram showing relation of three insulating phases to metallic rutile phase (gray) (b).
Transient reflectivity traces of microcrystal 27, $-\Delta R/R$, for low fluences where coherent phonon excitation is visible, with probe polarization parallel (purple) and perpendicular (gray) to the $c_R$ axis (c).
The initial pulse excitation is indicated by the red line. 
Inset scalebar is $5~\mu$m.
Fourier transform spectra of the reflectivity traces (d), showing phonon modes at 200 cm$^{-1}$ and 220 cm$^{-1}$.
Reflectivity at 3 ps for different probe polarizations (e). 
}
\label{Fig:Raman}
\end{figure}

The VO$_2$ micro-crystals, 
grown by vapor transport \cite{Guiton2005}, show distinct behavior from thin film samples, with an abrupt, first order transition in the absence of strain or doping. 
With their well defined, controllable strain state, it is possible to establish the strain-temperature phase diagrams for the material (see Fig.~\ref{Fig:Raman}b) \cite{Park2013,Atkin2012,Cao2010,Tselev2010}.
Peltier and Joule heating effects at metal-insulating domain interfaces have also been observed, probing the complex domain wall structure between the phases \cite{Favaloro2014}.
Microcrystals have been used for nano-mechanical devices \cite{Holsteen2014} and their domain structure and transition behavior can be readily tuned by the application of external mechanical stress \cite{Cao2009}.

The crystals have a well-defined size (Fig.~1d), crystallographic orientation, and thermal transition temperature, which we characterize with atomic force microscopy (e), micro-Raman spectroscopy (Fig.~2a), and scattering-scanning near-field optical microscopy ($s$-SNOM) \cite{Jones2010}.
The very low luminescence background and narrow linewidths in Raman spectroscopy indicate a low defect density, and the crystals are highly structurally homogeneous, free of defects and grain boundaries, indicating a single crystal structure with a flat surface (rms roughness $< 1.8~$nm), as seen from the atomic force microscopy image. 
They show sharp thermal phase transition behavior \cite{Wu2006}, without percolation behavior observed in thin film samples \cite{Qazilbash2007}. 
The shift of the $\omega_{\rm V-O}$ Raman mode 
allows us to estimate the amount of strain within the crystal due to substrate interactions or doping. 
Three possible insulating structures can be distinguished, monoclinic 1 or 2, or intermediate triclinic (M1, M2, T) \cite{Atkin2012}.
The combination of Raman with pump-probe microscopy allows us to confirm the high structural quality of the individual VO$_2$ micro-crystals and to systematically study the relationship of crystallographic orientation, insulating phase, and temperature with the photoinduced response dynamics.
IR nano-imaging using $s$-SNOM allows us to monitor the growth and evolution of metallic and insulating domains through the thermal transition with 10 nm spatial resolution.

\section{Results}
For low pump fluences, the transient reflectivity response shows an initial electronic excitation, due to the above-gap excitation of the pump, followed by relaxation on a picosecond time scale, as seen in Fig.~2c.
Probing at 800 nm (1.5 eV), the response is expected to be dominated by electrons in the d$_{\parallel}$ bands close to the Fermi level \cite{Verleur1968,Wall2013,Veenendaal2013}. 
The modulations in the reflectivity signal indicate the excitation of coherent phonons  in the insulating phase \cite{Zeiger1992}. 

An orientational anisotropy is evident in the relaxation and coherent phonon behavior, with faster decay
 and more prominent oscillations for pump polarization perpendicular to the crystallographic $c$-axis (gray line), compared to parallel polarization (purple line). 
The Fourier transform phonon spectrum, shown in Fig. 2d, reveals an even more pronounced anisotropy. 
For polarization parallel to the $c_R$ axis, only one low energy phonon peak is resolved, at $\sim200$~cm$^{-1}$ (6 THz).
For perpendicular polarization, both phonon modes emerge, with the second at approximately 225~cm$^{-1}$ (6.7 THz). 
Similarly, the reflectivity change at 3 ps also shows an angular anisotropy, with a cos$^2\theta$ dependence with angle $\theta$ of pump polarization with respect to the $c_R$ axis (Fig. 2e).

At higher fluences, the coherent phonon response is reduced, eventually vanishing for fluences sufficiently high to drive the microcrystal through the photoinduced insulator-metal transition (Fig.~3).
The persistence of the reflectivity change (up to microseconds) indicates that we have induced a quasi-stable metal-like state.
Without a simultaneous structural probe, we cannot determine the crystallographic structure during this period and whether the excited state corresponds to the postulated monoclinic metallic state \cite{Kim2006,Tao2012,Morrison2014}.
The inset shows the reflectivity $-\Delta R/R$ at 1 ps as a function of fluence, in order to derive the threshold fluence F$_{\rm th}$. 
F$_{\rm th}$ for different microcrystals varies between 2 mJ/cm$^2$ and 6 mJ/cm$^2$, which we attribute to variable coupling to the substrate. 
These values are close to or slightly lower than those observed in thin films, which range from approximately 5.5 mJ/cm$^2$ up to $> 15$ mJ/cm$^2$ \cite{Wall2013, Cavalleri2004, Cocker2012, Morrison2014}, 
due to stronger substrate coupling.

In order to quantify the transition dynamics and relate to the physical characteristics of the different microcrystals, we fit the above-threshold transient reflectivity behavior to $-\Delta R/R(t) = I(t) \bigotimes f(t)$, where $f(t)$ is the response function of the microcrystal, modeled as a bi-exponential function:
\begin{equation}
f(t) = -\frac{R_\infty}{1+a}(e^{-t/\tau_f}+ a e^{-t/\tau_s})+R_\infty.
\end{equation}
This is convolved with $I(t)$, the transient intensity of the pulse as determined by frequency resolved optical gating (FROG), which accounts for the time resolution and the step-like excitation at $t= 0$ without a priori assumptions about pulse duration or shape (see supplement for more details on the convolution procedure and results on fits).
Here $\tau_f$ describes the initial, ultrashort timescale transition behavior, and is constrained to sub-1~ps. 
$\tau_s$ captures the long timescale behavior, with 
$2~\rm{ps} < \tau_s < 10~\rm{ps}$, $R_0$ describes the magnitude of the reflectivity change, and the parameter $a$ sets the relative scaling of the fast and slow exponential terms.
We find that the contribution from slow dynamics is generally small (i.e. $a \ll 1$) and therefore we focus here on the short timescale dynamics captured by $t_f$. 
By fully characterizing the pulse using FROG, we can more accurately resolve the sub-50 fs dynamics.

Representative fits of the transient reflectivity response for microcrystal 13
for a range of above-threshold fluences are shown in Fig. 3b (black dashed lines), with $-\Delta R/R$ normalized for clarity.
For the $< 500$ fs range shown here, the response is dominated by the $\tau_f$ term.
We observe three distinct characteristics in the ultrafast initial response dynamics.
First, for the lowest fluence investigated we see a transition time of $\tau_f = 40 \pm8$ fs, shorter than previously observed dynamics in any VO$_2$ sample, as discussed further below. 
Second, we observe a dramatic, up to three-fold {\em increase} in transition time with increasing fluence (inset),  in contrast to a decrease observed in previous work on polycrystalline films \cite{Wall2013}.
Third, we see the transition time $\tau_f$ decrease with increasing sample temperature
(Fig. 5c). 
During the measurement we used sufficiently low laser repetition rates ($<100$ kHz) to allow complete relaxation to the initial insulating state, thus minimizing effects due to hysteresis that can occur close to the transition temperature. 
Previous observations note a decrease in threshold fluence when the initial temperature is increased \cite{Kubler2007,Pashkin2011}.
Through the combination with Raman spectroscopy we can correlate this behavior with the structural changes of the insulating phase on heating, as has been observed previously \cite{Atkin2012}. 
The pump-probe measurement at $T = 352$~K shows only a very small change in transient reflectivity with very fast dynamics, due to the presence of metallic stripe domains from thermal excitation, before the arrival of the pump pulse.

\begin{figure}
\includegraphics[width = \columnwidth]{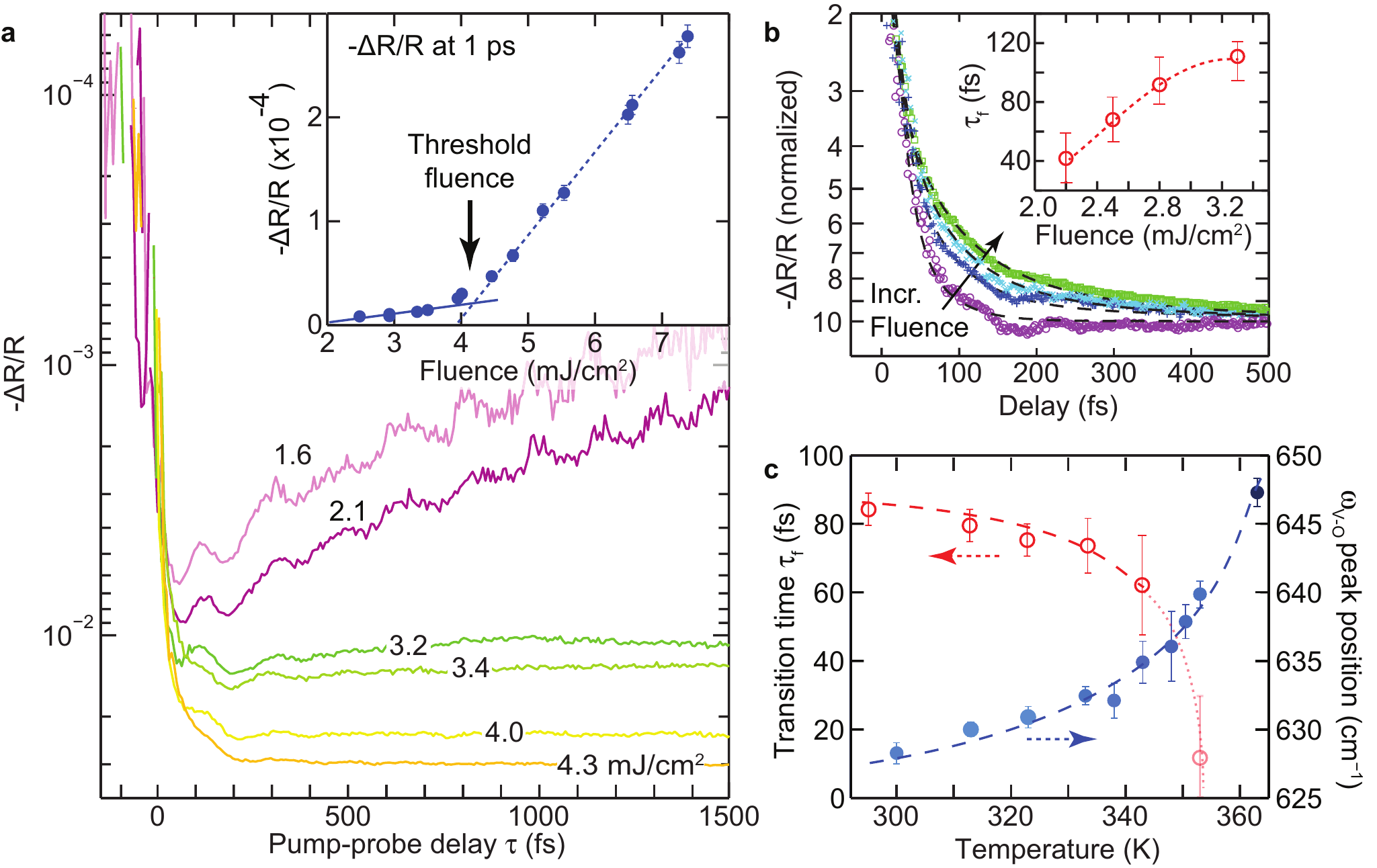}
\caption{
Fluence dependence of transient reflectivity $-\Delta R/R$ for microcrystal 26(a).
Change in short timescale dynamics with fluence for $F > F_{\rm th}$ (b), on microcrystal 13.
Black lines show fits to exponential recovery behavior, with extracted transition time $\tau_f$ shown in the inset. 
Variation of transition time with temperature, measured at 3.3 mJ/cm$^2$, showing a decrease in $\tau_f$ with increasing temperature, measured for microcrystal 26 (c).
This is correlated with a change in insulating structure, as monitored through the $\omega_{\rm V-O}$ Raman mode.
Blue and red dashed lines are guides to the eye.
}
\label{Fig:scatter}
\end{figure}

The data shown in Fig.~3 are for two different microcrystals; Fig.~3(a,c) show fluence and temperature dependence for microcrystal 26, respectively, and Fig.~3(b) shows data for microcrystal 13. 
The trends of an increase in $\tau_f$ with fluence and a decrease with temperature are seen for all measured microcrystals, though extracted transition times vary between crystals. 
The transition times $\tau_f$ obtained from all measurements on 20 microcrystals of different sizes, and for selected different initial temperatures and range of fluences are summarized in Fig.~4.
The values are shown as a function of the measured $\omega_{\rm V-O}$ phonon frequency from Raman spectroscopy, as a proxy for the different insulating phases of the crystals (indicated by green-blue color bar for the M1, T, and M2 phases).
The error bars are based on the uncertainty of the fit of eq. (1) in the $y$-direction, and a Lorentzian fit to the V-O phonon Raman line in the $x$-direction. 
Different microcrystal widths are indicated by the size of the data symbols (see legend), from less than 5$~\mu$m to greater than 15$~\mu$m. 
The fluences used are all above the threshold fluence for the specific microcrystal, with values indicated using false color. 
The values vary from 40 fs to 200 fs. 
Notably, the average value of $\tau_f$ over all microcrystals is found to be $\overline{\tau_{f}} = 80~\pm~25$ with $\overline{\tau_{f}}^{-1} = \sum_i \tau_i^{-1}$ (blue circle). 
This value is in striking agreement with transition times $\tau_{\rm TF}$ from thin film studies \cite{Cavalleri2004b}.

\begin{figure}
\includegraphics[width = \columnwidth]{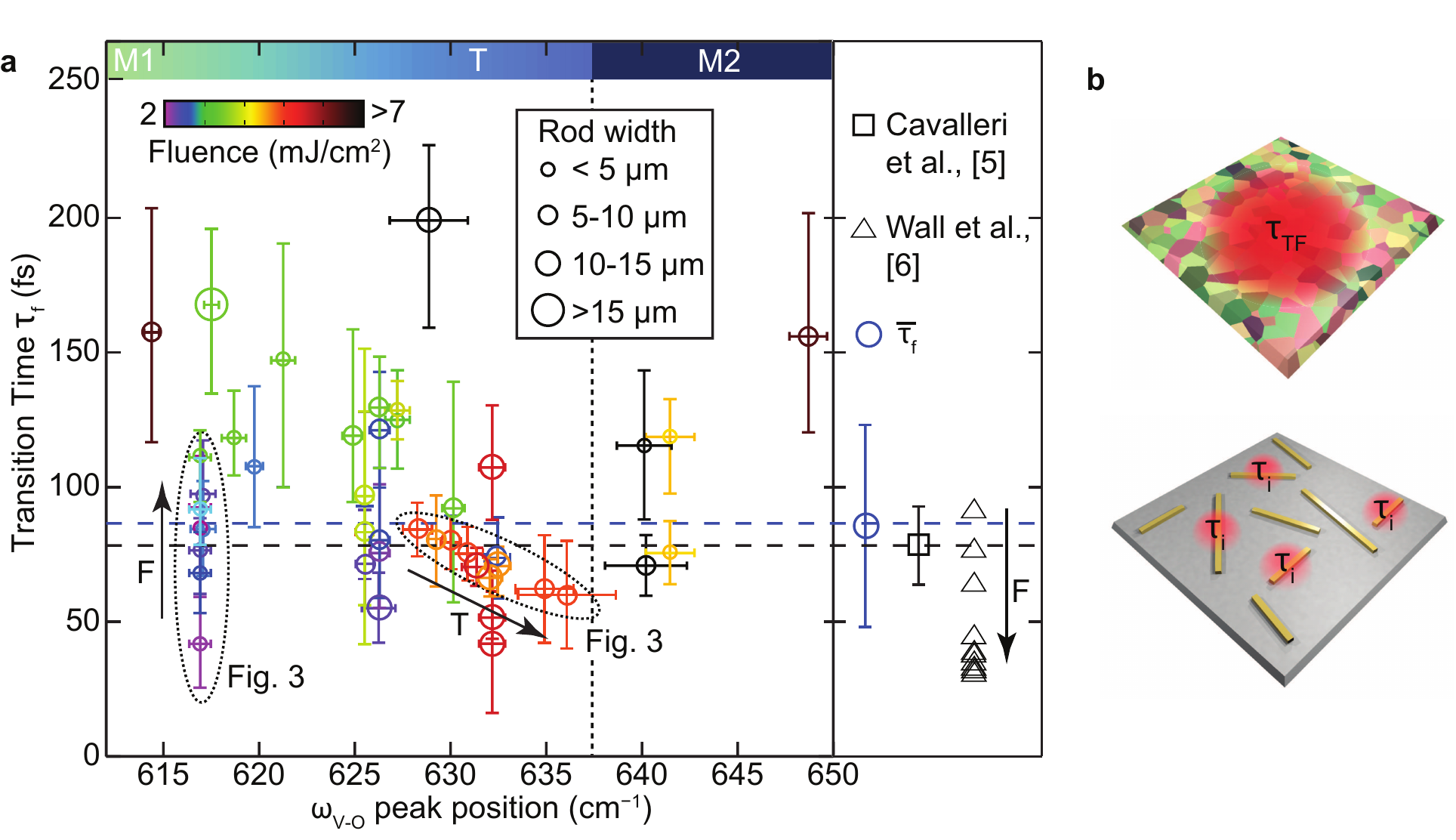}
\caption{Transition time for ultrafast photoinduced transition, $\tau_f$, plotted against initial ambient insulating phase, as given by the position of the Raman mode $\omega_{\rm V-O}$.
Colorbar at top shows insulating phases. 
Size of the data points reflects the microcrystal width, from less than 5 $\mu$m, to more than 15 $\mu$m. 
Color indicates the fluence at which the transition time $\tau_f$ was measured, from 2 mJ/cm$^2$ (orange) to $> 6$ mJ/cm$^2$ (black). 
The measured transition times vary from 40 fs to 200 fs, with no clear correlation with insulating phase.
Dashed ovals indicate data sets also shown in Fig. 3.
The blue dashed line and circle show the average transition time based on the ensemble of measurements, of 80 $\pm$ 25 fs. 
The black dashed line and square show 75 fs, the limiting timescale observed in thin film samples by Cavalleri et al.\cite{Cavalleri2004b}.
The red-black triangles show fast time constants extracted from data by Wall et al. \cite{Wall2013}, where a strong decrease in time constant is observed with increasing fluence, in contrast the behavior observed here for single crystals. 
Schematic of thin film (top, $\tau_{\rm TF}$) vs single crystal (bottom, $\tau_i$) measurements of the transition time, where the average transition rate over all crystals provides a value close to the thin film value (b). 
}
\label{Fig:scatter}
\end{figure}

\begin{figure}
\includegraphics[width = 12 cm]{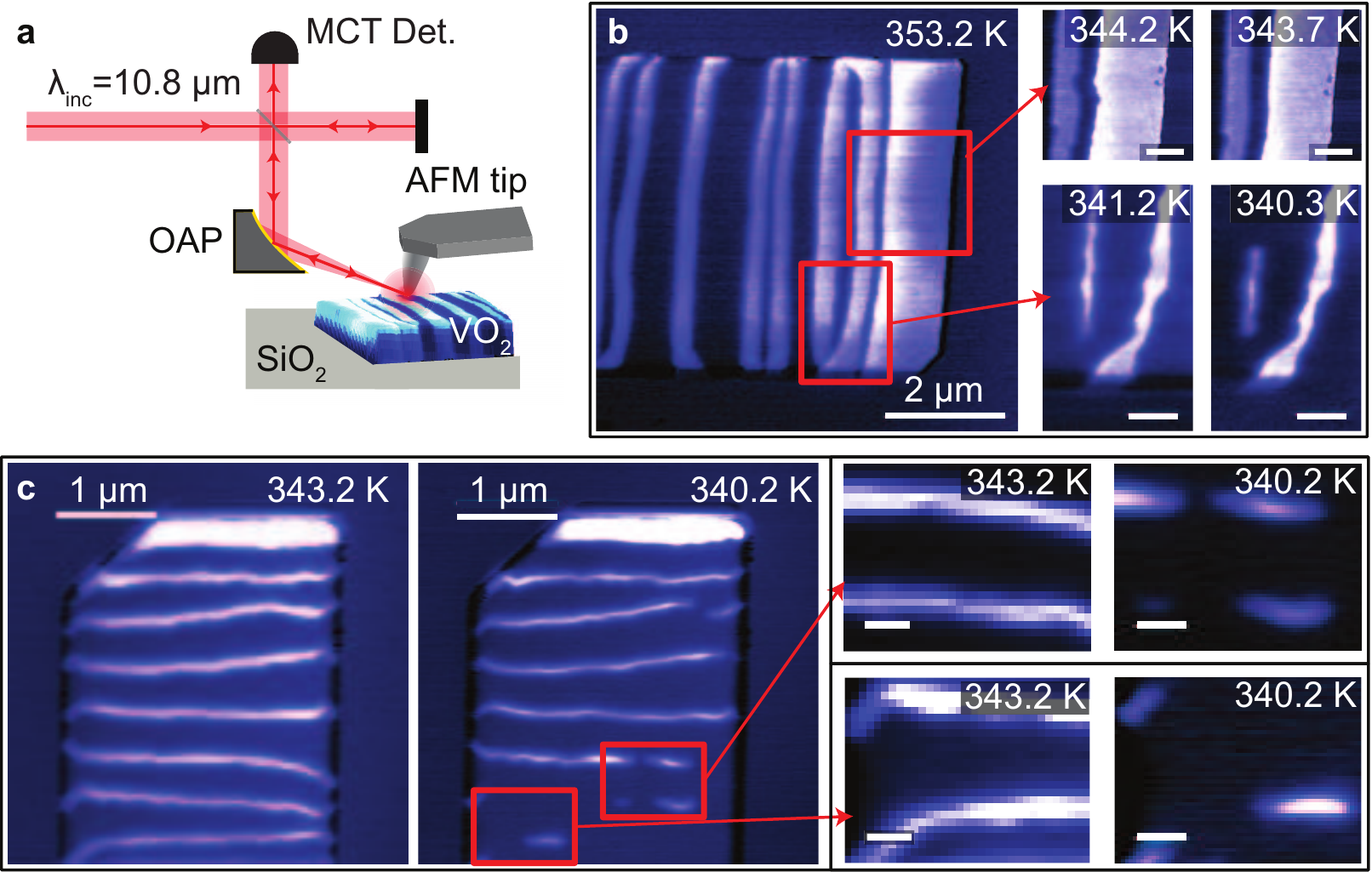}
\caption{Schematic of set-up for scattering scanning near-field optical microscopy ($s$-SNOM) imaging (a).
10.6$~\mu$m laser light is focused onto an AFM tip, and the scattered light is collected and detected using an MCT detector. 
When heating through the metal-insulator transition, the microcrystals form alternating metallic (gray) and insulating (dark blue) domains along the crystallographic $c$-axis in order to minimize substrate strain \cite{Jones2010} (b). 
On cooling, the metallic domains similarly break up along the $c$-axis direction, but then gradually narrow, forming meandering metallic lines, as shown in the inset to (b) and (c). 
The metallic domains then slowly break and form nanoscale metallic puddles ((c), insets).
The scale bars in the insets of (b) and (c) are $500$ nm and $200$ nm, respectively.
Over repeated cooling cycles, the domains break in different locations and at varying temperatures (see supplement). 
}
\label{Fig:sSNOM}
\end{figure}

The ultrafast pump-probe results indicate that the photoinduced IMT is highly inhomogeneous among the single crystallites. 
This inhomogeneity appears to be uncorrelated with structure, strain, temperature, or microcrystal size, with no consistent pattern of behavior observed even between crystals attached to the substrate (i.e., strained) or free. 
This suggests a sensitivity to moderate variations in the doping, stoichiometry, or defects.
These variations are small enough that they are not reflected in the lattice structure at the level detectable by few-wavenumber Raman shift, nor in the overall strain or temperature dependence. 
These variations between different crystals in turn may be spatially inhomogeneous and  lead to spatial variations in the IMT on the intra-crystalline level, within the individual single crystals. 
We take advantage of the complex phase behavior of microcrystals attached to the substrate, where the minimization of strain to accommodate the different thermal expansions of VO$_2$ and the silicon substrate leads to the formation of metallic stripe domains along the $c$-axis \cite{Jones2010} (Fig~5). 
The presence of these mesoscopic domains with the nanoscale spatial phase coexistence of both metallic and insulating regions  provides us with the sensitivity to resolve the effect of microscopic inhomogeneities on the thermal IMT, from the interplay of extrinsic strain and localized intrinsic defects and impurities at the domain walls.

Fig.~5a shows a schematic of $s$-SNOM nano-IR probing of the Drude dielectric response of VO$_2$ microcrystals on heating and cooling through the phase transition, with $\sim 10$~nm spatial resolution. 
Fig.~5b shows the formation of mesoscopic metallic stripe domains perpendicular to the $c$-axis on cooling from the metallic state, with straight domain walls and homogeneous behavior as the insulating states begin to form in the center of the microcrystal.
The insets show the gradual narrowing of the metallic domains upon cooling, associated with the emergence of complex, meandering domain walls with details very sensitive to temperature.
Once the metallic domains have narrowed sufficiently, they begin to break up in the direction perpendicular to the $c$-axis (insets, Fig.~5b and c). 
This inhomogeneous domain wall roughening and disappearance of the metallic domains on cooling is highly variable with repeated temperature cycling (see supplement).
This thermal behavior, with both reproducible and non-reproducible spatial features, supports a hypothesis of an electronically driven transition with both static and dynamic variations in the local properties, as discussed further below.

\section{Discussion}

In the following we discuss the implications of the above observations of the ultrafast photoinduced response dynamics and thermal spatial behavior of single VO$_2$ micro-crystals, especially with regard to the interpretation of previous experiments on polycrystalline thin films. 

Below threshold, we resolve the dynamical response of the two low-frequency modes in the coherent phonon spectrum, at approximately 200 and 225 cm$^{-1}$. 
These modes are of $A_{1g}$ symmetry, and are attributed to twisting of vanadium dimers, with their relative strength depending on the type of insulating phase and crystal orientation with respect to pump and probe polarization \cite{Schilbe2002,Zeiger1992}. 
Frequencies and line-widths are consistent with corresponding incoherent Raman scattering and track the different insulating phases, of M1, M2, and triclinic with their characteristic frequencies. 
Previous coherent phonon measurements have disagreed in whether one or two phonon modes are observed in the 200 cm$^{-1}$ range (6 THz vicinity), and in their precise frequencies \cite{Kim2006, Wall2012}.
These findings can now be reconciled given possible different insulating phases of the crystallites in the thin films.
The modes that are observed can thus depend on the number and relative orientation of the ensemble of crystallites probed in polycrystalline films.

Above the fluence threshold for the transition, we observe ultrafast dynamics on the tens of fs timescale, with a collapse in the gap in the insulating phase. 
In thin film measurements, the fluence-dependent behavior has been divided into three regimes:
below threshold, where coherent phonons are resolved; above threshold, where the system is driven into the metallic state but thermal effects are visible over longer timescales; and a saturation regime, where the magnitude of the transient reflectivity signal saturates and long timescale thermal behavior is no longer observed \cite{Wall2013}.
In contrast, our fluence dependent measurements on single micro-crystals show only two distinct regimes (Fig. 3): the below threshold regime where coherent phonons are observed and the reflectivity relaxes over picoseconds, and the above threshold regime with emergence of the quasi-stable metallic phase with no long timescale thermal behavior.
In the saturation regime, in general the rapid initial change and persistence of the change in reflectivity indicates that the entire probed volume experiences an ultrafast photoinduced transition to a metal-like state. 
The micro-crystal thickness does not appear to affect the transition time.
Since the crystals are thinner than the penetration depth of 800 nm light of approximately 180 nm, a homogeneous excitation of the microcrystal can be assumed. 
In contrast, a polycrystalline ensemble consists of crystallites with different threshold fluences and different dynamics.  
The ensemble measurement will then appear to be a superposition of multiple timescales, requiring a larger number of fitting parameters \cite{Wall2013}. 

Most notably we find that the initial insulating phase (M1, M2, or T) has no influence on the dynamics of the ultrafast transition. 
Furthermore, crystals with apparently identical lattice structure as concluded from identical Raman spectra reveal different photoinduced transition time scales.
This suggests that the emergence of the metallic phase in the photoinduced IMT is not a lattice related effect and the variations in the IMT dynamics point to an electronic delocalization transition. 

With the numerous observed transition times below the 75 fs half period of the $\omega_{V1,V2}$ phonon modes, the fastest being $40 \pm 8$ fs, across a range of fluences, we deduce that the 150 fs timescale for breaking the V-V dimer bonds is not relevant as a rate limiting step for the the formation of the metallic state of the photoinduced transition, as originally proposed by Cavalleri {\it et al.} \cite{Cavalleri2004b,Baum2007}. 
Our range of timescale values is similar to those reported in Wall {\it et al.} \cite{Wall2013}, but their results are not directly comparable to ours since they are based on a model with additional time constants and fitting parameters.

The electronic origin of the dynamics is supported by the decrease in transition time observed on heating the microcrystals.
While the free energy change with the increase in temperature is small \cite{Pouget1974} compared to the energy of the pump pulse, 
the change in dynamics we observe is substantial.
The insulating phase also changes with increasing temperature, following the progression M1-T-M2 (as shown in in Fig.~3), but the structural change in itself appears to have no effect on the transition time, as discussed for the comparison between microcrystals. 
However, these results suggest a finite response time for the transition \cite{Wegkamp2014}. 

The temperature-dependent decrease in transition time is an interesting counterpoint to the increase in transition time with increasing fluence. 
In contrast, the observations of Wall {\it et al.} indicate that the time constants in their model all decreased with increasing fluence \cite{Wall2013}.
Our results suggest a possible artifact of thin-film polycrystalline studies which probe an ensemble averaged response of variable numbers of crystallites, each with varying transition times and threshold fluences, however differences in the fitting functions preclude a direct comparison of extracted time constants.

Fluence behavior similar to that observed here has been previously observed in graphite \cite{Ishioka2008} and Cr-doped V$_2$O$_3$ \cite{Mansart2010}.
For VO$_2$, this could suggest that the higher fluences drive the system further out of equilibrium and lead to slower transition times to the metallic state. 
A possible mechanism would be non-equilibrium interband excitations, with variable fractional pump-induced occupation depending on the density of states and its variation with doping and impurities, to states supporting or opposing band structure collapse (Fig.~6b).
Saturation of states favorable to fast transitions could occur at higher fluences, and would lead to slower dynamics with increasing fluence due to increased excitation of states opposing band structure collapse.

In conclusion, even on the individual crystal level, for nominally homogeneous single crystals, we observe inhomogeneous behavior. 
We conclude therefore that in thin films the intrinsic dynamics can be masked by the inhomogeneous distribution and complicating extrinsic interactions among crystallites, with different sizes, orientations, and strain (Fig.~4b).
Our results therefore call for more precise characterization of VO$_2$ samples in order to gain a better understanding of the {\em intrinsic} response. 
{\em s}-SNOM imaging of the thermal transition also shows inter- and intra-crystal inhomogeneity: on heating, the nucleation and growth of the metallic domains is highly reproducible, but the break-up on these metallic domains on cooling can vary over repeated thermal cycles (see supplement).
The formation of metallic and insulating domains during the transition arises due to strain reduction, but minimization of free energy in a homogeneous system, considering bulk thermodynamic energy, strain energy, domain wall energy, twinning, and insulating lattice structure, should favor straight domain walls rather than the meandering structure we resolve in Fig.~5b and c. 
Furthermore, the metallic puddles remaining after break-up of the stripe domains appear at varying positions and are extremely sensitive to temperature, indicating that the VO$_2$ microcrystals are in a highly dynamic state close to the critical temperature $T_C$.

\begin{figure}
\includegraphics[width = 11 cm]{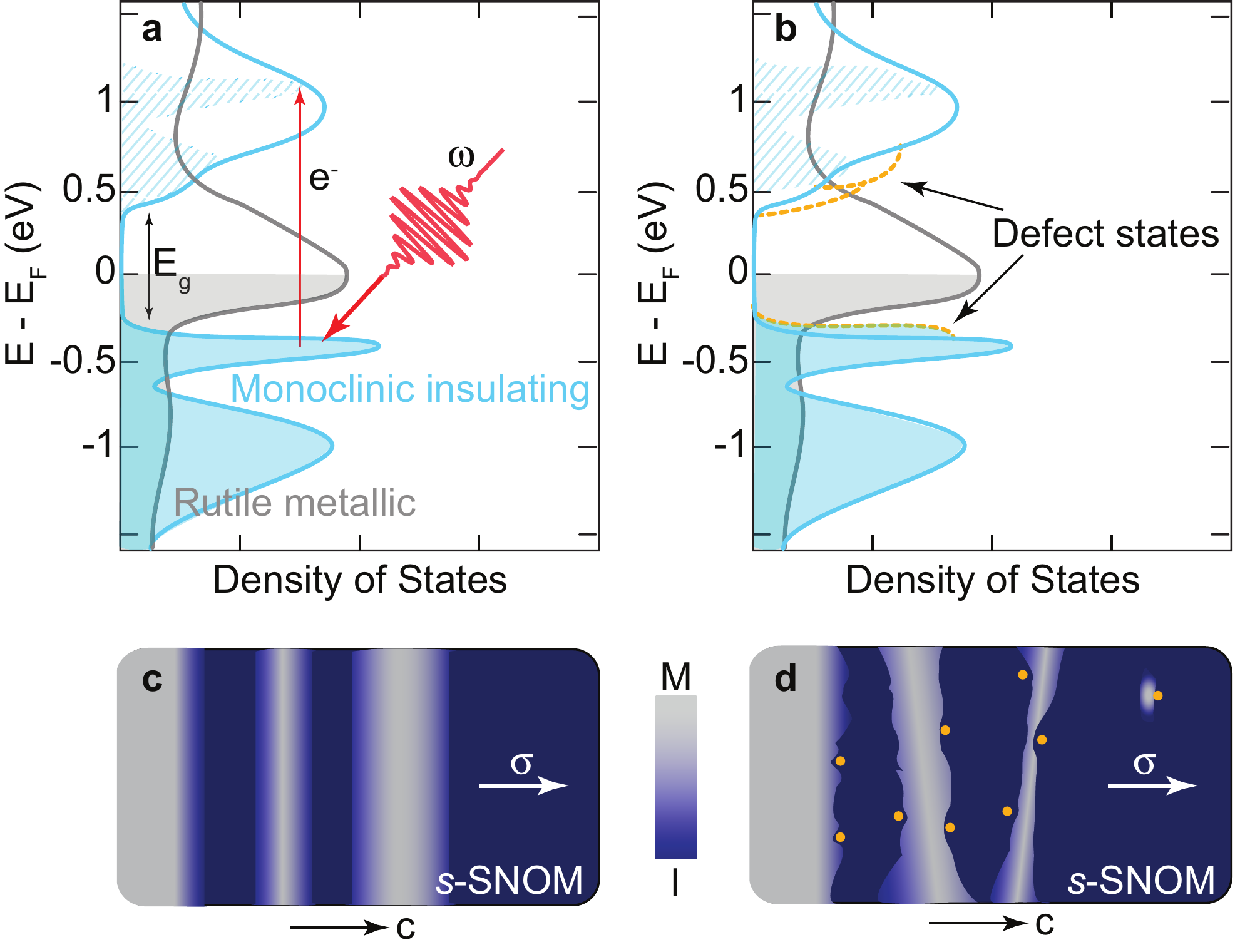}
\caption{Schematic representation showing the electronic band structure of VO$_2$ (a) and possible modifications from defects or impurities (yellow lines, b), which could alter the transition time of the ultrafast IMT and produce the variable dynamics observed.
Defects could also change the spatial arrangement of metallic and insulating domains in single crystals, producing a shift from straight domain walls, as predicted by theory (c), to more complicated structures (d). 
}
\label{Fig:sSNOM}
\end{figure}

Both the ultrafast photoinduced studies and thermal IMT $s$-SNOM therefore 
show a high degree of sensitivity to the effect of dopants and defects.
These could alter the density of available states and therefore the redistribution of holes and electrons due to the exciting pulse, and thereby change the rate of electron delocalization and subsequent bandgap collapse (see Fig.~6a,b). 
Similarly, defects and complex strain could disrupt the free energy uniformity and produce complex domain topology in the thermal transition, as we resolve at domain walls for attached crystals through $s$-SNOM (Fig.~6c,d).
Further study of the broadband response of microcrystals could help to elucidate the properties of these defects and their effect on the mechanism of the the IMT. 

We reveal a wide variation in static and dynamic properties of apparently homogeneous, well-characterized single crystal sub-systems. 
These results raise the question of how to access the intrinsic response of VO$_2$, and that of strongly correlated electron materials more generally. 
The rich and diverse properties of these materials that can be induced and controlled through doping, strain, external fields, etc. may be more sensitive to disorder and impurities than previously expected.

\section{Experimental Methods}
\noindent
The single crystal vanadium dioxide microcrystals studied were grown by vapor phase transport on an oxidized silicon substrate \cite{Guiton2005}.
This produces rectangular microcrystals of varying sizes and orientations, as shown in Fig.~1d). 
Typically the rods have width 100 nm - 15 $\mu$m, and length up to 1 mm. 
The rutile $c$-axis ($c_R$) is along the length of the rod.

The photoinduced IMT in individual microcrystals is studied using a degenerate pump and probe transient reflectivity measurement, with experimental configuration shown in Fig.~1a. 
Excitation is provided by a regenerative amplifier Ti:S system (K\&M Labs, Wyvern), which produces $<50$~fs pulses at 800 nm, with a variable repetition rate from $10-350$ kHz. 
The 800 nm pump provides above band gap excitation, while the probe at this wavelength is dominated by the response of electrons in the d$_{\parallel}$ bands close to the Fermi level \cite{Wall2013}. 
A 5:95$\%$ beamsplitter separates the incident light into the probe and pump arms, with a delay of up to 100 ps introduced, with resolution 0.3 fs. 
The two beams are then recombined with a small spatial offset, passed through a dual frequency optical chopper, and focused onto the sample using an off-axis parabolic mirror with probe focus size $\sim 15~\mu$m. 
This produces pump fluences of up to 10's of mJ/cm$^2$. 
The reflected light is collected again by the parabolic mirror, and the probe light is detected by the photodiode (New Focus, Nirvana 2007) with lock-in amplification in order to improve the signal-to-noise ratio. 
In addition, the pump and probe polarizations are orthogonal, and a polarizer is used to select the probe signal. 
Pulses are fully characterized with frequency resolved optical gating (FROG) (Fig. 1 b,c).

The samples are mounted on a resistive heater with a thermocouple to enable temperature control within $\pm~0.5$~K.
Atomic force microscopy (AFM) measurements are used to characterize the heights of the rods chosen for measurements.
From this, we find the rods have rectangular cross-sections with thicknesses of 25 ~-~200~nm (Fig.~1e).
Simultaneous with the AFM measurements, we perform scattering-scanning near-field optical microscopy ($s$-SNOM) with a CO$_2$ laser source to probe the changes in reflectivity at 10.6 $\mu$m as crystallites move through the IMT.
We see stripe domain formation due to the strain between the microcrystal and the substrate, shown in Fig.~5, as previously observed \cite{Jones2010,Liu2013}.

We characterize the crystallography of individual microcrystals through Raman spectroscopy, using a home-built microscope with HeNe laser excitation ($\lambda = 632.8$ nm) and a 0.8 NA objective (Olympus).
The Raman scattered light is detected by a spectrometer with 600 groove/mm grating and liquid nitrogen-cooled CCD after passing through a cut-off filter, enabling 2 cm$^{-1}$ spectral resolution. 
The three insulating structural phases, monoclinic M1, monoclinic M2, and triclinic T, can be distinguished principally through the position of the 610-650 cm$^{-1}$ Raman mode, denoted $\omega_{\rm V-O}$ (Fig. 2a) \cite{Atkin2012}. 
Microcrystals with $\omega_{\rm V-O}$ close to 650 cm$^{-1}$ are in the M2 phase, which can be produced by substrate strain or $\sim 2 \%$ Cr doping \cite{Jones2010,Atkin2012} .
For microcrystals in the M1 phase, $\omega_{\rm V-O} \sim 620$ cm$^{-1}$, and the amount of doping or impurity is less than $1\%$ \cite{Marini2008}.
In addition to this characterization of insulting phase, we observe a polarization dependence of the Raman modes.


\end{document}


\title{Supplement to: Inhomogeneity in the ultrafast insulator-to-metal transition dynamics of VO$_2$}

\author{Brian T. O'Callahan}
\thanks{These authors contributed equally to this work.}
\author{Joanna M. Atkin}
\thanks{These authors contributed equally to this work.}
\author{Andrew C. Jones}
\affiliation{Department of Physics, Department of Chemistry, and JILA, University of Colorado, Boulder, CO 80309}
\author{Jae Hyung Park}
\author{David Cobden}
\affiliation{Department of Physics, University of Washington, Seattle, WA 98195}
\author{Markus B. Raschke\footnote{Corresponding author: markus.raschke@colorado.edu}}
\affiliation{Department of Physics, Department of Chemistry, and JILA, University of Colorado, Boulder, CO 80309}

\date{\today}
\maketitle
\section{Details on fitting procedure and pulse deconvolution}
As mentioned in the main text, we perform frequency resolved optical gating (FROG) using a home-built apparatus to gain access to the transient phase $\phi(t)$ and amplitude $A(t)$ of the pulse. 
With the intensity profile $I(t) = A(t)^2$, we convolute the fitting function in the frequency domain to obtain $-\Delta R/R(t) = I(t) \bigotimes f(t)$ where $f(t)$ is a bi-exponential function representing the response function of the microcrystal:
\begin{equation}
f(t) = R_0(e^{-t/\tau_f}+ a e^{-t/\tau_s})
\end{equation}
where $\tau_f$ describes the initial, sub-1 ps timescale behavior, $\tau_s$ captures the long time scale behavior constrained to between 2 and 10 ps, and $a$ is the relative amplitude between these two exponentials. 
The convolution procedure simulates a step-like excitation at $t=0$ and incorporates the time resolution resulting from the finite duration of the pump and the probe pulse. 

\section{Tables of crystals}


\begin{longtable}{| p{.13\textwidth} |p{.2\textwidth} | p{.15\textwidth} |p{.10\textwidth} |p{.10\textwidth} |p{.10\textwidth} |p{.19\textwidth}|} 
\hline
    Crystal No. & Fluence (mJ/cm$^2$) & $\omega_{V-O}$ (cm$^{-1})$ & $\tau_f$ (fs) & $\tau_s$ (fs) & $a$ & Crystal size ($\mu$m) \\
		\hline
    1     & 3.3 & 617.5 & 168 & 3580 & 0.02 & 33.6 \\
    2     & 5.3 & 631.4 & 70 & 2000 & 0.03 & 12.3 \\
    3     & 4.5 & 629.3 & 80 & 2050 & 0.1 & 5.5 \\
    4     & 3.3 & 630.2 & 92 & 2500 & 0.3 & 7.3 \\
    5     & 4.5 & 632.1 & 66 & 2000 & 0.09 & 10.4 \\
    6     & 4.5 & 632.5 & 71 & 2500 & 0.01 & 11.3 \\
    7     & 3.3 & 621.3 & 147 & 2000 & 0.3 & 4.9 \\
    8     & 3.0 & 627.2 & 125   & 2860 & 0.4 & 4.8 \\
          & 3.7 & 627.2 & 128 & 2010 & 0.2 & 4.8 \\
    9    & 3.3 & 624.9 & 119   & 2000 & 0.3 & 6 \\
    10    & 3.3 & 618.7 & 118 & 2100 & 0.3 & 2.9 \\
    11    & 2.5 & 616.9 & 68 & 2000  & 0.1 & 3 \\
          & 3.3 & 616.9 & 111   & 2590 & 0.2 & 3 \\
    12    & 2.4 & 626.2 & 55 & 2020 & 0.2 & 10 \\
    13    & 2.2 & 616.9 & 42 & 8620 & 3E-6 & 3 \\
          & 2.4 & 616.9 & 77 & 5000  & 0.01 & 3 \\
          & 2.5 & 616.9 & 92 & 2000  & 0.01 & 3 \\
          & 2.8 & 616.9 & 92 & 2000 & 0.2 & 3 \\
    14    & 2.4 & 625.5 & 71 & 10000 & 0.01  & 9 \\
    15    & 2.3 & 626.3 & 76 & 9720 & 3E-5 & 9 \\
          & 2.5 & 626.3 & 121 & 2760 & 0.4 & 9 \\
          & 2.5 & 626.3 & 80 & 2010 & 0.05 & 9 \\
          & 3.3 & 626.3 & 130 & 2010 & 0.2 & 9 \\
    16    & 2.4 & 617.1 & 97 & 9940 & 0.009 & 2.3 \\
          & 2.6 & 617.1 & 84 & 9990 & 0.01 & 2.3 \\
    17    & 2.6 & 619.8 & 108 & 2000 & 0.2 & 4.8 \\
    18    & 3.6 & 625.5 & 97 & 2000 & 0.1 & 9 \\
          & 3.6 & 625.5 & 83 & 3600 & 0.3 & 9 \\
    19    & 7.0 & 614.4 & 157 & 2000 & 0.2 & 6 \\
    20    & 7.0 & 648.7 & 156   & 8250 & 0.002 & 6 \\
    21     & 5.3 & 632.2 & 42 & 9990 & 5E-7 & 11.5 \\
          & 5.3 & 632.2 & 107 & 2000  & 0.05 & 11.5 \\
          & 5.3 & 632.2 & 52 & 6840 & 2E-7 & 11.5 \\
    22    & 8.2 & 640.2 & 71 & 10000 & 0.04 & 7.5 \\
    23    & 8.2 & 628.9 & 199   & 5860 & 0.01 & 10 \\
    24    & 4.1 & 641.5 & 76 & 6710 & 5E-6 & 2.5 \\
          & 4.1 & 641.5 & 119 & 2000 & 0.03 & 2.5 \\
    25    & 8.2 & 640.1 & 115 & 2010 & 0.2 & 4.5 \\
		26    & 4.5 & 628 & 87.03 & 2000 & 0.1 &10 \\
		27    & 0.8 & 636 &   &  &  &7.2 \\
\hline
  \caption{Fit parameters shown in Fig.~4 of main text.}
\end{longtable}%

\begin{longtable}{| p{.13\textwidth} |p{.2\textwidth} | p{.15\textwidth} |p{.10\textwidth} |p{.10\textwidth} |p{.10\textwidth} |p{.19\textwidth}|} 
\hline
    Crystal No. & Fluence (mJ/cm$^2$) & $\omega_{V-O}$ (cm$^{-1})$ & $\tau_f$ (fs) & $\tau_s$ (fs) & $a$ & Temperature (K) \\
		\hline
    27    & 3.3 & 628.3 & 84.2 & 2000 & 0.2 & 297 \\
         & 3.3 & 630.0 & 79.5 & 3150 & 0.45 & 313 \\
         & 3.3 & 630.9 & 75.3 & 2010 & 0.15 & 323 \\
         & 3.3 & 632.5 & 73.7 & 9090 & 1.4E-5 & 334 \\
         & 3.3 & 634.9 & 62.1 & 9920 & 0.012 & 343 \\
         &  3.3 & 640.0 & 7.6 & 5000 & 0.01 & 353 \\

\hline
\caption{Table of fit parameters shown in Fig. 3c of main text.}
\end{longtable}%

\section{Additional temperature dependence data for microcrystal 26}
Here we present additional data on temperature dependence of microcrystal 26 regarding Fig. 3c of the main text. Raman spectra are shown in Fig.~\ref{RamanHys}a,b for heating and cooling cycles, respectively.
Metallic domains appear at 348 K and the microcrystal is fully metallic above 385 K.
Additionally, we observe the evolution of the insulating phase through the T phase with increasing temperature and the remaining insulating stripes transition to M2 at 363 K. 
This is indicated by the shift in the $\omega_{\rm V-O}$ peak position displayed in Fig.~\ref{RamanHys}c.
A $~10$ K hysteresis in the peak position is observed for the T-M2 transition.
Table 2 correlates the extracted $\omega_{\rm V-O}$ peak position with the ultrafast dynamics.

\begin{figure}[!b]
\includegraphics[width = \columnwidth]{../figures/Raman-hysteresis.eps}
\caption{Raman spectra of crystal 26 on heating (a) and cooling (b) through the thermal transition. Metallic stripes appear at 348 K and the microcrystal is fully metallic above 385 K. Extracted $\omega_{\rm V-O}$ peak position showing the evolution of the insulating phase through the T phase upon heating. At 363 K, the remaining insulating stripes transition to the M2 phase.}
\label{RamanHys}
\end{figure}

\section{Further {\it s}-SNOM imaging}
Here we provide further nanoscale images showing the irreproducible break up pattern of metallic puddles upon cooling through the thermal transition. 
Fig.~\ref{fig:sSNOM} shows {\it s}-SNOM images during repeated cooling cycles. 
The red circles indicate areas where there are deviations in the spatial pattern of the remaining metallic domains.
The irreproducibility illustrate the highly dynamical state of the microcrystals near the critical temperature $T_C$ which is extremely sensitive to spatial variations in the chemical potential and perturbations in the electronic structure. 
This highly dynamical state is in part a result of the free energy degeneracy of the rutile metallic and the two monoclinic insulating structures at $T_C$ \cite{Park2013}.
As the crystal is cooled from $T_C$, the degeneracy is broken and free energies begin to deviate, favoring the insulating phases.
The insulating areas that first nucleate are observed to be in the M2 phase \cite{Jones2010} due to the substrate-induced strain, which is predicted to have a structural free energy contribution closer to the rutile structure $\Delta H_{M1 \rightarrow R} > \Delta H_{M2 \rightarrow R}$ \cite{Pouget1974} and is therefore more sensitive to fine spatial variations in the free energy landscape. 
This 

\begin{figure}
\includegraphics[width = \columnwidth]{../figures/supplement3.eps}
\caption{{\it s}-SNOM imaging during repeated cooling cycles showing irreproducible break-up pattern of metallic domains on a VO$_2$ microcrystal. Scale bars are 500 nm. Scans from the first cooling sequence (top row) show a meandering pair of domains on the left that break around $T = 344.0$ K at nearly the same locations in the different cycles, but exhibit different domain patterns at $T = 343.2$ K (red circles) that cannot be explained by hysteresis effects since they begin and end with roughly the same domain pattern. This also may not be explained by difference in temperature since they are qualitatively different domain patterns that do not match a corresponding temperature in the other cycle.}
\label{fig:sSNOM}
\end{figure}

Fig.~\ref{fig:sSNOM} shows several thin metallic domain patterns during two repeated cooling cycles imaged with {\it s}-SNOM. 
The domains on the left break at roughly the same location at $T = 344.0$ K, but they show different domain patterns upon cooling to $T = 343.2$ K, as highlighted in the red circles.
Since the domain patterns begin and end in qualitatively the same pattern, this cannot be explained by hysteresis effects which could result in a slightly different domain pattern at high-T or at lower temperature.
The pattern discrepancy in the scan does not match scans at different temperatures for the other cooling cycle, so this is not a result of slightly different temperature variations between the temperature sweeps.
These metastable nanoscale metallic domains are observed to relax to a completely insulating state over temperature steps as small as $\Delta T = 0.1$ K, indicating their dynamic nature.